\documentstyle[psfig]{res}
\begin{document}

\title{%
DEPENDENCE OF STAR FORMATION RATE ON OVERDENSITY}

\author{Kentaro Nagamine \\
{\it    Physics Department, Princeton University, Princeton, NJ, 08544, USA,
nagamine@astro.princeton.edu}\\
Renyue Cen, Jeremiah P. Ostriker\\
{\it Princeton University Observatory, Princeton, NJ, 08544, USA, \\
(cen, jpo)@astro.princeton.edu}}

\maketitle

\section*{Abstract}

We use a large-scale $\Lambda$CDM hydrodynamical simulation to 
assess the dependence of the cosmic Star Formation Rate (SFR) 
on overdensity of luminosity.

\section{Introduction}

The star formation rate (SFR) is the key measure that connects the 
structure formation in the universe and the actual observable light that is 
emitted by the galaxies. The plot of SFR as a function of redshift, known 
as the `Madau Plot', depicts the evolution of the formation rate of 
stars in galaxies.

Using a large-scale hydrodynamical simulation, it was shown by 
Blanton et al. (1999) and Nagamine et al. (1999) that the gas which falls 
into gravitational potential-wells gets shock-heated, and  
the higher overdensity regions become no longer preferred sites for 
galaxy formation as the temperature increases towards the present.

Here we present the effects of this phenomena from different 
directions by analysing the SFR as a function of overdensity of
luminosity. Details of the simulation can be found in the above 
two papers.

\section{Star Formation Rate vs. Overdensity}

The SFR is a direct output of our hydrodynamical simulation.
Figure 1 shows the SFR divided into the quartiles of the 
light-overdensity distribution in V-band at z=0. The V-band 
luminosity of each stellar particle in the simulation 
was obtained using the latest isochrone synthesis model
of GISSEL99 (Bruzual \& Charlot 1999) with the metallicity 
of each particle taken into account.

It is clearly seen that the stellar particles in higher 
overdensity regions form earlier in time than those in lower
overdensity regions. This is due to the effect which we
described in the Introduction; the gas in high overdensity regions
is shock-heated as it falls into the potential-well, thus 
further star formation is prohibited 
by the high temperature of the gas. Hence the steeper
turn-off of the SFR between z=1 and 0.
In lower overdensity regions, the gas is less heated
than the higher overdensity regions and the 
peak of the SFR is at lower redshift.

Observationally, this is well known as the `morphology-density 
relation' (eg., Dressler 1980). In clusters of galaxies,
the old non-star-forming early-type galaxies dominate, 
whereas in the field, star forming late-type galaxies
are seen more often. 
Also, the recent observations of galaxies in voids by 
Grogin \& Geller (1999, 2000) find that the 
galaxies in the lowest density environments show stronger star
formation than those in higher density region, which is 
consistent with what we find in Figure 1.

\begin{figure}[t]
\centerline{\psfig{figure=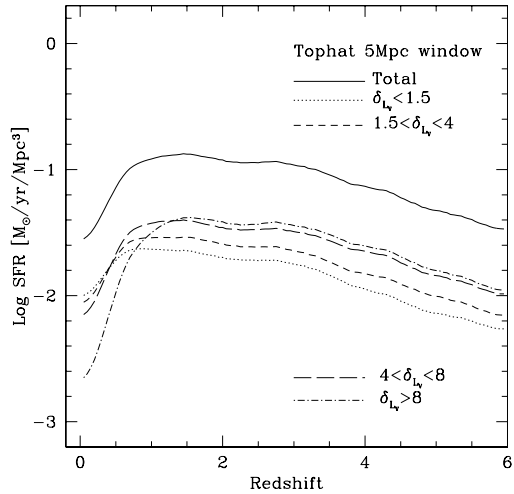,width=7cm,height=7cm,angle=0}}
\caption{SFR as a function of redshift, divided into the quartiles of the 
light-overdensity distribution in V-band at z=0.}
\end{figure}

\section{References}

\re
1.      Blanton, M. et.al., 1999, submitted to {\it ApJ}, astro-ph/9903165.
\re
2.	Bruzual, G. A. and Charlot, S., 1999, in preparation.
\re
3.	Dressler, A., 1980, {\it ApJ}, {\bf 236}, 351.
\re
4.      Grogin, N. A. and Geller, M. J., 1999, {\it AJ} in press, astro-ph/9910073.
\re
5.	Grogin, N. A. and Geller, M. J., 2000, {\it AJ} in press, astro-ph/9910096.
\re
6.      Nagamine, K., Cen, R., J. P. Ostriker, 1999, submitted to {\it ApJ}, astro-ph/9902372.

\end{document}